\documentclass{mem}
\usepackage{natbib}\usepackage{txfonts}\usepackage{balance}
\usepackage{graphicx}
\usepackage[a4paper]{hyperref}
\idline{75}{282}

\newcommand{\micron}{$\mu$m}
\newcommand{\kms}{km\,s$^{-1}$}

\newcommand{\Myr}{$M_{\odot}$\,yr$^{-1}$}
\newcommand{\msun}{$M_{\odot}$}
\newcommand{\lsun}{$L_{\odot}$}
\newcommand{\water}{H$_2$O}

\begin{document}
\def\teff{$T\rm_{eff }$}
\def\kms{$\mathrm {km s}^{-1}$}

\title{
AGB and post-AGB stars
}

   \subtitle{}

\author{
Dieter\,Engels\inst{1} 
          }

  \offprints{D. Engels}

\institute{ Hamburger Sternwarte,
Gojenbergsweg 112,
D-21029 Hamburg, Germany\\
\email{dengels@hs.uni-hamburg.de}
}

\authorrunning{Engels }

\titlerunning{AGB and post-AGB stars}

\abstract{ Intermediate mass stars (1-8 \msun) evolve along the
  Asymptotic Giant Branch after completion of hydrogen and helium core
  burning. At the tip they lose for several ten to hundred thousand
  years copious amounts of mass and exhibit various forms of
  variability, as for example large-amplitude variations with periods
  of one to five years.  In the oxygen-rich circumstellar envelopes
  powerful OH, \water\ and SiO masers may operate. Part of the AGB
  population is converted to carbon stars. During the latest phases of
  AGB evolution the mass loss rates approach several $10^{-5}$ \Myr,
  so that the stars become invisible in the optical and in part in the
  near infrared.
  
  On departure from the AGB a fundamental transition in the mass loss
  process is taking place changing from a spherically symmetric
  outflow on the AGB to axi-symmetric or point-symmetric
  geometries. While the mass loss rates decrease to $\approx10^{-8}$
  \Myr, the velocities are strongly increasing.  Stars are now in
  their "post-AGB" phase.  Mounting evidence is gathered that during
  the very latest phase of AGB evolution and during the post-AGB
  phase, evolution proceeds on very short timescales, which in extreme
  cases is comparable to the working life of an astronomer. 

} 
\maketitle{}

\begin{figure}[t!]
\resizebox{\hsize}{!}{\includegraphics[clip=true]{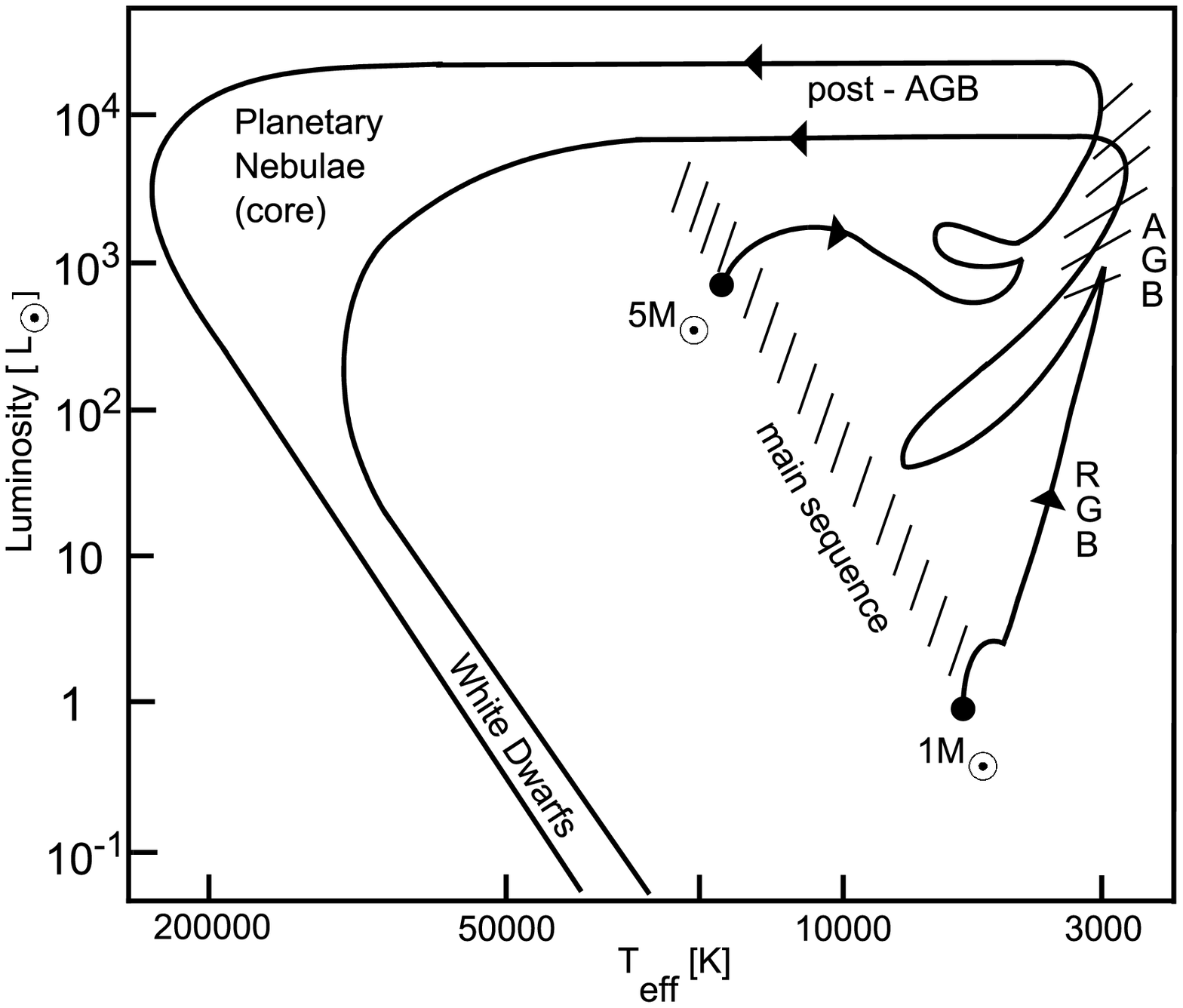}}
\caption{\footnotesize Schematic evolutionary tracks of stars with
main-sequence masses of 1 and 5 \msun\ and solar composition.
RGB = Red Giant Branch, AGB = Asymptotic Giant Branch.
}
\label{agb}
\end{figure}

\section{Introduction}
The time a star spends during its evolution on the Asymptotic Giant
Branch (AGB) is only a tiny fraction of $\le$1\% of its total
lifetime. Nevertheless this evolutionary phase sets the framework for
the later development into {\it stellar end products} like planetary
nebulae and white dwarfs. AGB stars are cool red giants and reach
their maximum luminosity during the AGB phase, making them easily
observable beacons in the sky. Strong infrared radiation, and in
part large-amplitude variability and masers allow their
efficient identification. The processes involved in the departure
from the AGB are still badly understood, because the stars 
are dust-enshrouded and inaccessible to direct view.
As post-AGB stars,
they increase their temperature at almost constant luminosity, until
they are able to ionize the remnant shell, creating one of the most
fascinating phenomena in the sky: planetary nebulae.

The purpose of the current review is an introduction to the non-expert
reader, covering the main observational properties of AGB and post-AGB
stars and the current understanding of the evolutionary
processes. References to further reading are given at the appropriate
places. A comprehensive coverage of the field is given by the book 
{\it Asymptotic Giant Branch stars} edited by \citet{Habing04}.

\section{Evolution on the AGB}
Stars enter the AGB when energy production in a double shell structure
begins, which is made up of an an inner burning He shell surrounding
an electron degenerated carbon-oxygen (C-O) core and an outer hydrogen
burning shell. The stars then already have left the main sequence and
have passed through the red giant branch (RGB), where energy had been
generated by hydrogen fusion in a shell around a core made up of
helium. The stars have completed also the core helium burning phase in
which the build-up of the C-O core had started.  On the AGB, double-shell
burning starts with a quiescent burning phase (the E(arly)-AGB) and
develops later in the thermal pulse phase (TP-AGB). The thermal
pulses are large energy releases generated by a flash-like burning of
the helium shell lasting several hundred years. Typical evolutionary
tracks in the Hertzsprung-Russell-Diagram (HRD) are shown in Fig. \ref{agb}.


The evolutionary time scales on the AGB are short compared to the main
sequence lifetimes. For 1--5 \msun\ stars the E-AGB phase lasts
$\approx$1--15 million years, compared to the time spent on the main
sequence of $\approx$0.1--12 billion years.  In the TP-AGB phase
stars remain another several hundred thousand years. Thus the duration
of the AGB phase is 0.5\% or less of the main sequence
lifetime (\citealt{Vassiliadis93}, \citealt{Bloecker95a}).  On the
TP-AGB the stars lose most of their mass and they finally leave the
AGB, when virtually only the naked C-O core is left over.  The time
scales derived from evolutionary models strongly depend on the adopted
mass loss law.
A critical review discussing the current understanding of the
mechanisms leading to the strong mass loss in AGB stars and of its
parameterization is discussed by \citet{Willson00}.
 
\section{Observational characteristics}
\subsection{Luminosities, masses, and chemistry}
Observationally the luminosities of AGB stars are confined to the range
$-3.6 \la M_{bol} \la -7.1$ ($2200 \la L/L_{\odot} \la 55\,000$) and they
have main-sequence masses between $\approx$0.85 and $\approx$8 \msun.
The lower limits are set by AGB stars occurring in globular clusters,
while the upper luminosity limit is observed for AGB stars in the 
Magellanic Clouds and is set by theory due to the (not strictly valid) 
core-luminosity relation, and the 
Chandrasekhar 1.4 \msun\ mass limit of white dwarfs \citep{Lattanzio04}.
The upper limit in mass is set by the observation of white dwarfs in open
clusters with turn-off masses $\la$8 \msun\  \citep{Koester96}.

Spectroscopically AGB stars can be divided into oxygen-rich types (spectral
type M), whose spectra are dominated by TiO bands, and carbon-rich types
(spectral type C), where bands of C$_2$ and CN dominate the
spectra. The difference is due to the carbon-oxygen ratio in the
atmospheres being C/O$<$1 for M-stars and vice versa for
C-stars. Transition objects with C/O$\approx$1 are known as
S-stars. As normal stars have C/O$<$1, carbon stars must be enriched
with extra carbon. According to AGB models this extra carbon is
brought to the surface by convection during the {\it third
dredge-up}. This process occurs after a thermal pulse, when the
convection zone is able to penetrate the hydrogen burning shell and
allows to dredge-up freshly processed matter \citep{Lattanzio04}.
A comprehensive review on carbon stars was given by \citet{Wallerstein98}.

\subsection{Variability}
As soon as the star enters the TP-AGB the stellar emission becomes
variable. It is thought that the light variations are caused by radial
pulsations and additional opacity variations in the stellar
atmospheres \citep{Reid02}. The most conspicuous variables having
large amplitudes are the optically prominent Mira variables and the
optically obscured infrared bright OH/IR stars, although irregular and
semiregular (SR) pulsators with small amplitudes are more
frequent. The periods of Mira variables are of the order $\approx$1
year and the visual amplitudes are large ($>$2.5 mag), while the periods of
OH/IR stars are between 2 and 5 years and the amplitudes in the near
infrared can reach a few magnitudes \citep{Engels83}. The light curves of
many Miras are sampled frequently due to the efforts of amateur
observers (see www.aavso.org), but this is not the case for OH/IR
stars because of the need of infrared or radio equipment.

\subsection{Mass loss, circumstellar shells } 
The energy generating core region is surrounded by a tenuous and
extended envelope in which mixing occurs by convection. The size of
the envelope may easily match earth's orbit.The outer layers are
bounded only weakly by gravitation so that the stars lose mass
already on the E-AGB at rates ($\la 10^{-7}$ \Myr) ten orders of
magnitude higher than that of the sun. 

The stellar pulsation on the TP-AGB extends the outer layers of the
envelopes to cooler regions facilitating the formation of molecules
and dust. Once dust has formed, radiation pressure is able to push it
away dragging the gas with it. Mass loss rates increase dramatically
reaching up to $10^{-4}$ \Myr\ and the outflows reach typical
velocities of $v_E = 10-15$ \kms. AGB CSEs have been reviewed by
\citet{Habing96}.

\begin{figure}[h] 
  \resizebox{\hsize}{!}{\includegraphics*[clip=true,angle=-90]{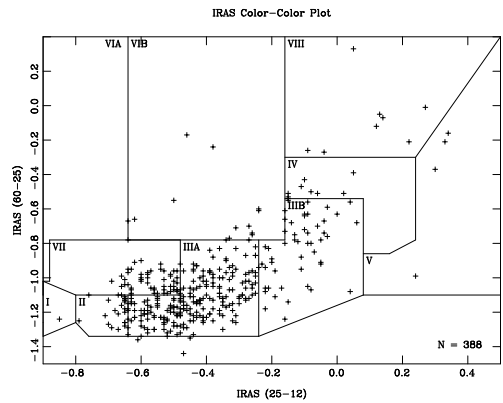}}
\caption{ \footnotesize IRAS color-color plot with the location of the
O-rich AGB stars from the 'Arecibo sample' \citep{Lewis94}. The labeled
regions were introduced by \citet{Veen88}. Regions I-IV contain mostly
O-rich stars, VI and VII mostly C-rich stars, and V and VIII many
post-AGB stars. }
\label{color_plot}
\end{figure}

The development of high mass loss rates leads to the formation of a
circumstellar gas and dust shell (circumstellar envelope, CSE) with
increasing optical depths. At rates $\ga 10^{-6}$ \Myr\ the stars
disappear optically. The stellar emission is absorbed by the dust and
re-emitted in the infrared. The obscured AGB stars were originally
discovered by OH maser surveys and subsequent identifications in the
infrared (``OH/IR stars''). Larger populations were later found by the
IRAS All-Sky Survey at 10--100 \micron, applying color selection
criteria. They divide IRAS color-color plots into regions
(Fig. \ref{color_plot}), in which AGB stars can be separated
efficiently from other objects. The distribution of OH/IR stars in
Fig.  \ref{color_plot} has the form of a sequence demonstrating
the wide range of optical depths involved. The blue end of the sequence
is made up by stars with optically thin CSEs (SR-, Mira variables), while
the objects in the red part of the diagram have optically thick CSEs
(OH/IR stars).

\begin{figure}[h] 
  \resizebox{\hsize}{!}{\includegraphics*[clip=true]{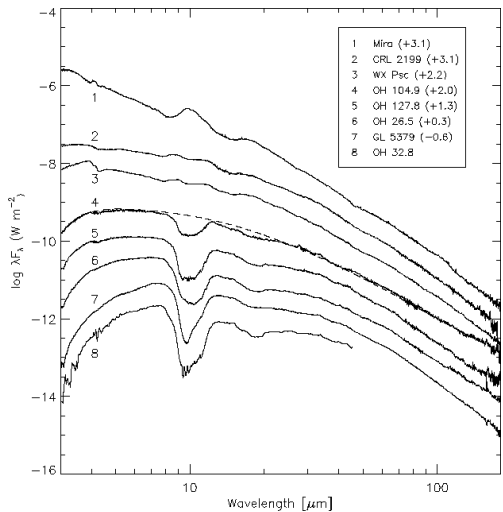}}
\caption{ \footnotesize ISO spectra of the 6--200\micron\ region of several 
O-rich AGB stars. The spectra are shifted to order them in terms of 
10\micron\ optical depth, which correlates roughly with IRAS colors and
mass loss rates (from \citet{Sylvester99}) }
\label{sed}
\end{figure}

The dichotomy between M- and C-stars is observed also in the infrared
as the chemistry of their CSEs differ according to the C/O-ratio of
the outflowing matter. As carbon is locked for C/O$>$1 almost
completely in the CO molecule, only oxygen-rich stars are able to form
oxygen-bearing dust.  This dust is mainly made of amorphous silicates
which exhibit a very strong feature at 9.7\micron\ in emission for
optically thin and in absorption for thick CSEs (Fig. \ref{sed}). In
contrast, the dominant dust species contained in carbon star CSEs is
graphite.  A classification even of obscured AGB stars discovered by
IRAS was therefore possible using the dust signatures found in the
spectra taken in the 10\micron\ range by the IRAS Low Resolution
Spectrometer (LRS).  The knowledge of the dust constituents has
tremendously increased with the detailed spectroscopy of AGB stars in the
mid-infrared by the Infrared Space Observatory (ISO)
(\citealt{Molster05}, \citealt{Blommaert05}).

\subsection{Masers}
\begin{figure}[h] 
 \resizebox{\hsize}{!}{\includegraphics[clip=true]{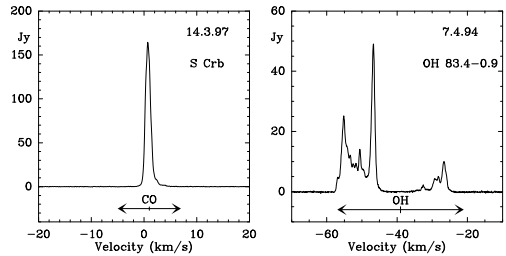}}
\caption{
\footnotesize
\water\ maser spectra taken with the Effelsberg radio telescope. The
Mira variable S\,Crb
displays a single maser feature close to the stellar radial velocity given
by the midpoint of the velocity range of the thermal CO emission.
The OH/IR star OH\,$83.4-0.9$ shows a double-peaked profile with a
velocity range almost identical as for OH maser emission.
}
\label{maser}
\end{figure}

The CSEs, extending to sizes $10^{16}-10^{17}$ cm ($>$1000 AU),
provide ample space for velocity coherent outflow required for
powerful maser emission. The emission always involves molecules
containing oxygen. The most prominent are the OH masers at 1612 and
1665/1667 MHz, the \water\ maser at 22 GHz and the SiO masers at 43
and 86 GHz. Typical maser spectra are shown in Fig. \ref{maser}.
Extensive reviews on masers in CSEs have been written by \citet{Cohen89}, 
and \citet{Elitzur92a}.

Maser surveys have shown that at low mass
loss rates SiO and \water\ masers are more frequently detected than OH
masers, while the detection rate of 1612 MHz OH masers is high at
higher mass loss rates. 
Also the maser profiles show systematics depending on mass loss rates
although they are often strongly variable in intensity. The 1612 MHz
OH masers are located at radial distances $\ge 10^{16}$ cm, and
propagate radially. Their profiles are double-peaked with the
blue-shifted peak coming from the approaching front side of the shell
and the red-shifted peak coming from the receding back side. The
double-peaked profile immediately tells the expansion velocity of the
shell (half the velocity difference between the peaks) and the radial
velocity of the star (midpoint of the OH maser velocity interval). In the
mm-wavelength range this information can be obtained also from the
profiles of thermal CO emission.

Because of their higher excitation temperature \water\ and SiO masers
are located more inside. The SiO masers were found to be located close
to the star inside the dust formation zone, while the \water\ masers
are located between the SiO and OH maser zones. At lower mass loss
rates these masers preferentially propagate in tangential directions,
displaying maser features close to the stellar radial velocity (S
Crb). In OH/IR stars the \water\ maser is located so far away from the
star that radial amplification as in OH masers takes place: the
typical double-peaked profile is displayed (OH\,83.4--0.9).

Dedicated radio interferometers like the VLBA and MERLIN are nowadays
used to map the distribution of the maser clouds around the star and
to measure proper motions.  Also the strengths and distributions of
magnetic fields in the CSEs can be determined by polarization
measurements of the masers (Vlemmings, these proceedings).  The most
spectacular result so far is the monitoring of the 43 GHz SiO maser of
the Mira TX\,Cam over a full pulsation period \citep{Diamond03}. The
observations revealed with biweekly time resolution the gas motions in
the near circumstellar environment of this star.  Together with the
new possibilities for high-spatial resolution near-infrared imaging
(see Wittkowski, these proceedings) the radio imaging of masers will
allow in the coming years studies of the processes leading to the high
mass loss rates observed in AGB stars in unprecedented detail.

\subsection{Samples}
TP-AGB stars may be unambiguously identified by detection of lines of
technetium in their optical spectra. Tc has only radioactive isotopes
and must have been dredged-up recently. However, in general secondary
criteria as the variability, the infrared excess, or the presence of
masers are used as selection criteria for AGB star samples.

They can be obtained from the General Catalogue of Variable stars
\citep{Samus04} by extracting stars classified as Miras or SR variables
\citep{Kharchenko02}. Also the carbon stars in the catalogue of
\citet{Stephenson89} are mostly AGB stars. Statistically well described
samples are easier to obtain from the IRAS Point Source Catalog applying color
criteria. For example, Lewis and collaborators \citep{Lewis94} provided a flux
limited sample of several hundred AGB stars ('Arecibo sample'). A sample of
AGB stars within 1\,kpc of the sun was compiled by Jura \& Kleinmann (see
\citet{Olivier01} for a recent discussion). Samples in other galaxies can be
obtained using the upper luminosity limit of 2500 \lsun\ valid for all RGB
stars \citep{Habing96} as lower selection threshold.  The only contamination
expected is from a few red supergiants.

\begin{figure}[t] 
  \resizebox{\hsize}{!}{\includegraphics*{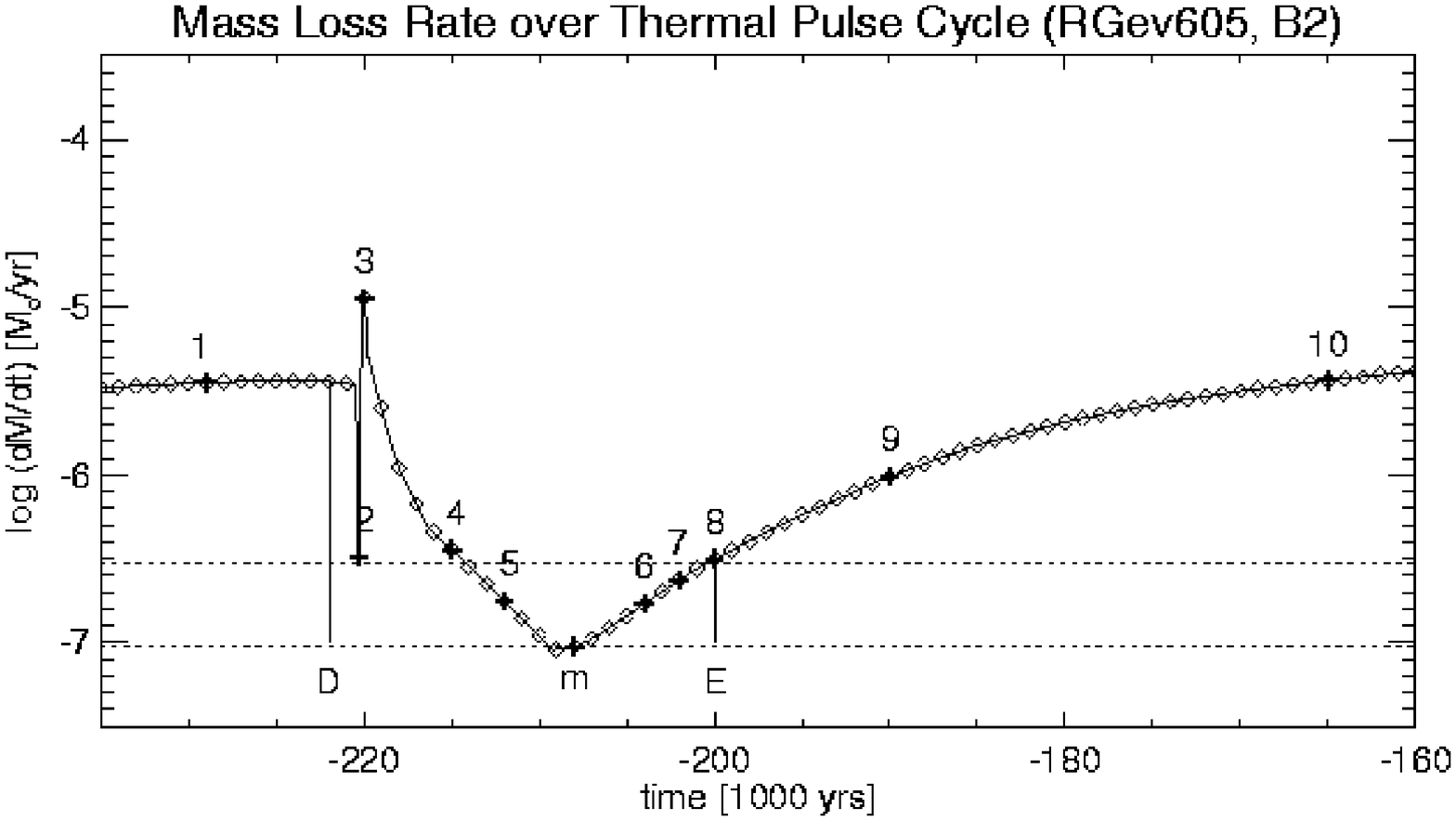}}
  \resizebox{\hsize}{!}{\includegraphics*{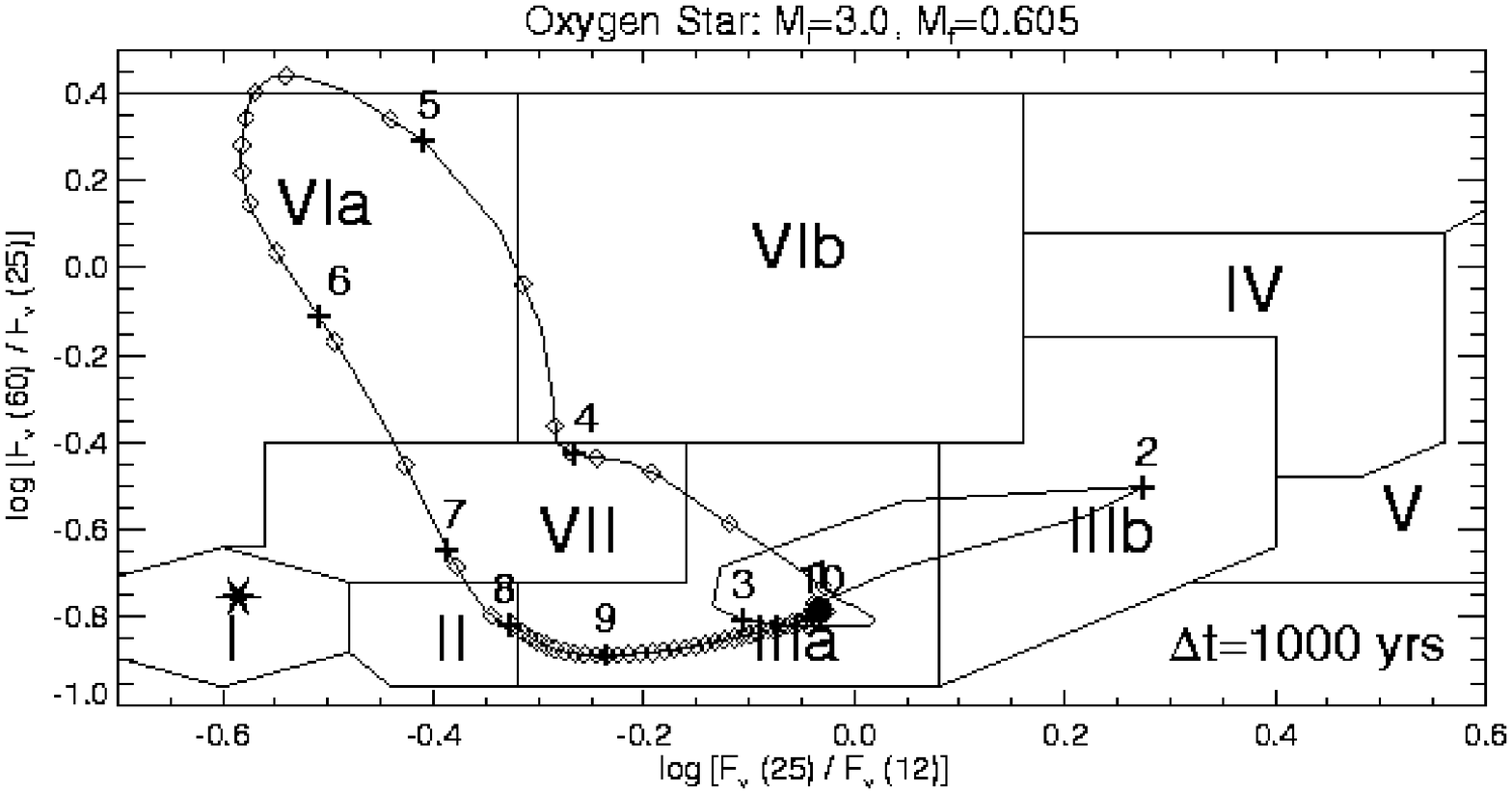}}
\caption{ \footnotesize 
Variations of the mass loss rate for one thermal pulse cycle and the
correspondent location of the star in the IRAS color-color diagram
of a 3 \msun\ model.  The separation between two adjacent
diamonds corresponds to 1000 years.  Labels ``1'' to ``10'' are used
as reference points (from \citet{Steffen98}).  
}
\label{closeup}
\end{figure}
 
\section{The departure from the  AGB }
AGB evolution ends when mass loss has removed the envelope leaving
behind the \mbox{C-O} core. Intuitively the sequence
SR$\longrightarrow$Mira $\longrightarrow$OH/IR star of increasing
mass loss rates and periods may be seen as an evolutionary sequence
along the AGB to higher luminosities regardless of initial mass
\citep{Bedijn87}. However, in response to the occurrence of thermal
pulses, variations in luminosity, periods, and mass loss rates are
expected and stars may cycle several times through the sequence
SR$\longrightarrow$Mira$\longrightarrow$OH/IR star. As example Figure
\ref{closeup} shows variations of the mass loss rate during a single
thermal pulse cycle for an 3 \msun\ model \citep{Steffen98}. As a result
the spectral energy distributions (SED) and the correspondent
synthetic IRAS colors will vary as well. The movement in the IRAS
color-color diagram is in the form of 'loops' (Fig. \ref{closeup}),
with their exact locations depending on chemistry and main-sequence
mass. The galactic height distribution in addition is not uniform
across the IRAS color-color diagram. Jim\'{e}nez-Esteban (these
proceedings) finds scale heights of 300--500 pc in the blue part and
100--200 pc in the red part of the color sequence, implying mass
segregation.  Only the more massive AGB stars may actually develop
into an OH/IR star before leaving the AGB.

Part of the stars will be converted into carbon stars and leave the O-rich
color sequence in the IRAS color color diagram. They may originate mainly from
stars in the mass range $\approx$2--4 \msun, because the lower mass stars may
not have sufficient thermal pulses and subsequent dredge-ups to invert the
original C/O$<$1 ratio and in the higher mass stars the carbon might be
quickly converted into nitrogen due to the so-called 'hot bottom burning'
\citep{Groenewegen04}.

Thus, stars will end their AGB evolution with different chemistries of their 
residual envelopes, and with different optical depths of their CSEs. This
depends not only on their main-sequence mass, but also on the exact
moment of mass exhaustion in the envelope during the thermal pulse cycle.
The departure from the AGB is marked by the cessation of large amplitude 
variability and by a remnant CSE, which is detached. The stars move
for a brief period (few $10^3$ years) into the extremely red parts
of the IRAS color-color diagram ([25--12]$>$0.0 in Fig. \ref{color_plot}).
Candidates are the 'non-variable OH/IR stars', in which the outermost
OH maser zone is still undisturbed, but the \water\ masers respond to
decreasing densities in the inner CSE \citep{Engels02}.

\begin{figure}[h] 
  \resizebox{\hsize}{!}{\includegraphics*[clip=true,angle=-90]{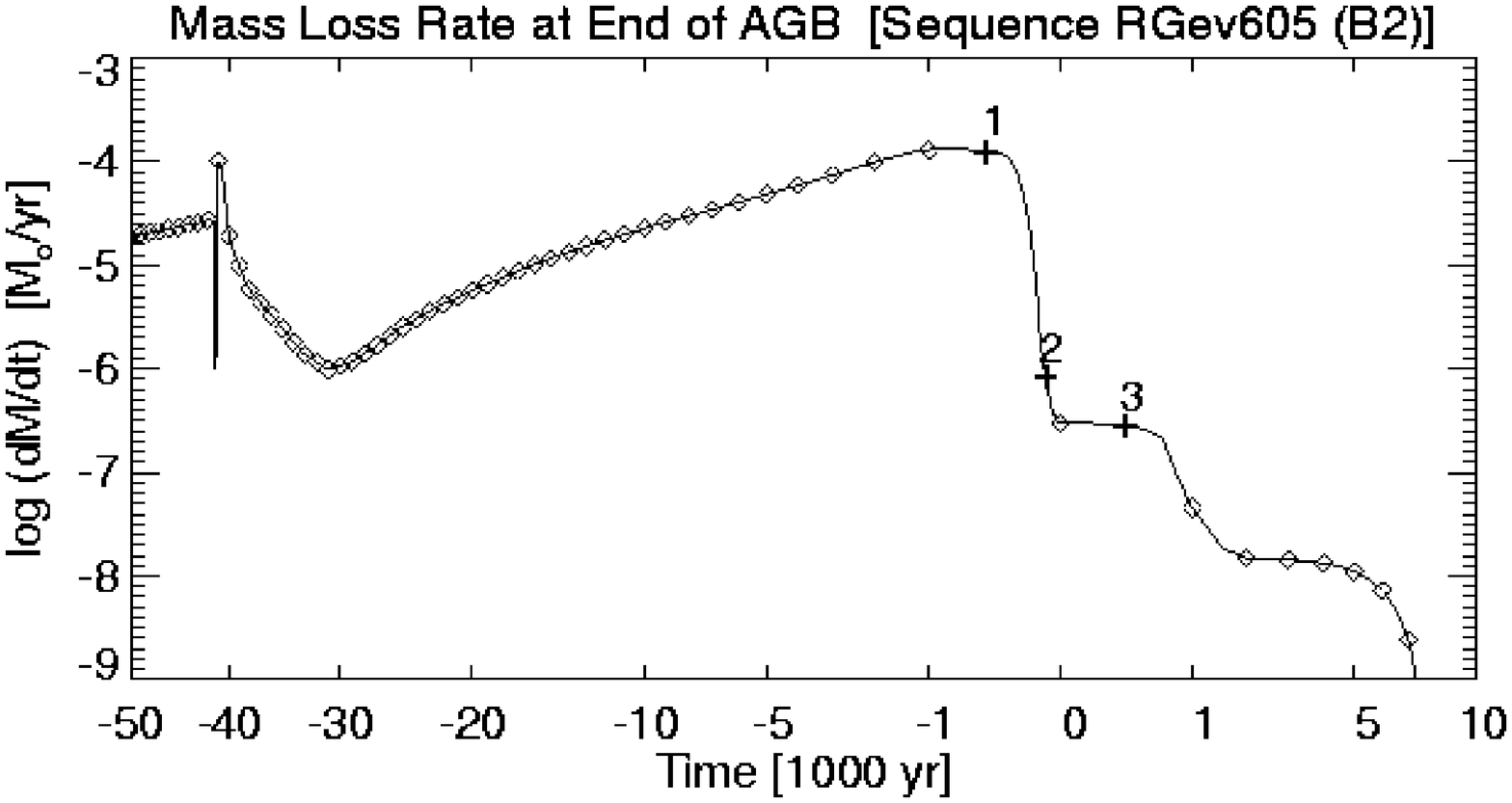}}
  \resizebox{\hsize}{!}{\includegraphics*[clip=true]{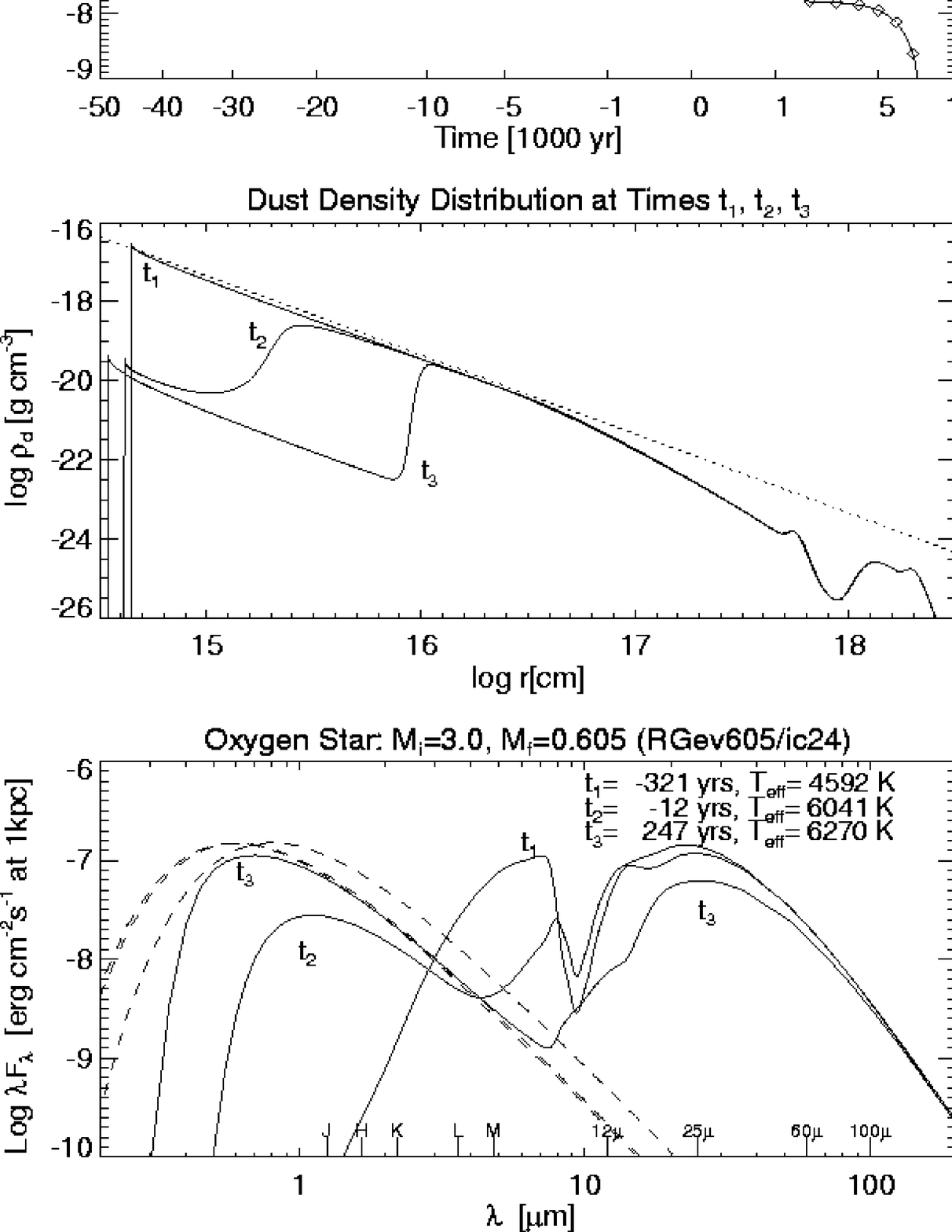}}
\caption{ \footnotesize 
  Development of the mass loss rate after the last thermal pulse (top) and
  development of a double-peaked spectral energy distribution (bottom) in
  response to the drop of the mass loss rate between time stamps t$_1$ and
  t$_3$  (from \citet{Steffen98}).  
}
\label{post_agb}
\end{figure}
 
\section{Post-AGB stars}
\subsection{Evolution}
As shown in Figure \ref{post_agb} the OH/IR star like heavily obscured SED at
the tip of the AGB will change on short timescales ($\approx500$ years for the
3 \msun\ model of \citet{Steffen98}) to a double-peaked distribution, in which
the central star shines through the now diluted CSE. Parallel to the mass loss
reduction, the star (the former core) now begins to contract and heat up. In
the HRD (Fig. \ref{agb}), it develops almost horizontally with constant
luminosity to higher temperatures. If the temperature reaches 25\,000 K the
radiation is able to ionize the remnant CSE, well visible then as planetary
nebula. Surprisingly, the post-AGB mass loss rate was not found to be zero
albeit small ($\approx10^{-8}$ \msun). Furthermore the winds speeds are much
higher than on the AGB reaching $10^3$ \kms.

The time scales that models predict for the post-AGB phase range
between $10^1$ and $10^5$ years depending inversely on the core mass,
the treatment of the mass loss reduction process, and the mass loss
history on the AGB \citep{Bloecker95b}. This implies that not all
post-AGB stars will eventually become a planetary nebula. For low core
masses the remnant CSEs might have dissipated before the stars are able
to ionize the shell. For high core masses the temperature rise is so
fast that the CSEs may still be opaque and emission from the ionized
inner rim is trapped.

\subsection{Observational characteristics}
Considerable numbers of post-AGB stars were discovered using the IRAS
All-Sky Survey to search for optical bright objects with strong
mid-infrared excesses, e.g. stars with a double-peaked SED.  Other
searches focused on the red parts of the IRAS color color diagram,
providing optically fainter post-AGB stars. A recent compilation
contained 220 post-AGB stars and candidates (see Szczerba et al. in
\citet{Szczerba01}). Statistically well defined samples are still
lacking and new approaches to find new post-AGB stars are clearly
warranted (Richards, these proceedings). Optical spectroscopy of
post-AGB stars demonstrated that in accordance with the wide range of
temperatures covered by the evolutionary tracks, all spectral types
between B and K are present.  Although lines of elements dredged-up
during the AGB phase are detected, the chemical composition of the
photospheres show a larger diversity as anticipated. The phase of the
thermal pulse cycle when the star leaves the AGB may determine the
chemical composition of the remaining envelope. But also the
membership in a binary system seem to have more influence on
observable properties than on the AGB, where only a few members in
binary systems are known.
 
\begin{figure}[t!]
\resizebox{\hsize}{!}{\includegraphics[clip=true]{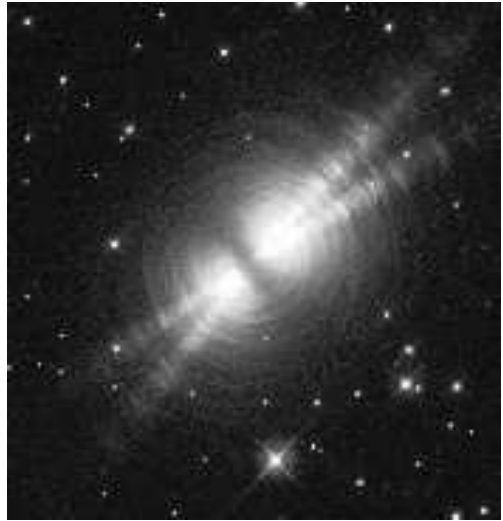}}
\caption{\footnotesize Proto-Planetary Nebula AFGL 2688 (Egg Nebula).
Credit: NASA and The Hubble Heritage Team (STScI/AURA)}
\label{egg}
\end{figure}

\subsection{Mass loss and nebula shapes}
Planetary nebulae and protoplanetary nebulae (PPN) show a large diversity of
morphologies \citep{Balick02}. The most remarkable structural change,
even for many very young PPN, is the loss of the radial
symmetry prevalent for the mass loss on the AGB. A case in point is
the Egg nebula (Fig. \ref{egg}), which displays a bipolar nebula
morphology, created by polar outflows originating from the center of
an optically thick dusty disk. The nebula is surrounded by spherically
symmetric concentric arcs, which evidently are associated with the
remnant AGB shell.  The gaps between the arcs infer modulations of the
AGB mass loss rates on time scales of a few hundred years, which
neither fit to the thermal pulse nor to the envelope pulsation time
scales.

The velocities of the bipolar (sometimes even multipolar) outflows have been
mapped in about a dozen PPN in CO or in the \water\ or OH masers revealing
velocities up to $10^3$ \kms. The time scales implied for the age of these
outflows are several hundred to a few thousand years, implying that the
departure from the AGB happened only recently (see also Gomez, these
proceedings).  After having passed the planetary nebula stage stars descend
the white dwarf cooling track. However, some may experience a very late
thermal pulse bringing them back to the AGB ('born-again AGB stars'). One case
is Sakurai's object (V4334 Sgr) evolving with a fascinating pace on the
time-scale of years across the HRD (Eyres, these proceedings).

Research on post-AGB stars is a lively field and we still lack a coherent
picture, matching evolutionary theories with the bewildering diversity of
post-AGB star types. The accumulation of knowledge proceeds fast and is
reflected by recent publications of \citet{Kwok00}, \citet{Szczerba01},
\citet{VanWinckel03} and \citet{Waelkens04}.


\section{Brief outlook}
Research on AGB and post-AGB stars has advanced strongly in the last 25 years,
beginning with the discovery of thousands of new objects by IRAS. Key open
questions encompass the primary mass loss mechanism, the influence of thermal
pulses on observational properties and the origin for the transition of radial
symmetric mass loss on the AGB to axi- or point-symmetric high-velocity
outflows in the post-AGB phase. The understanding of the key physics for the
mass loss process is of great importance, because both the AGB evolution
models as well as the models to synthesize the AGB population in foreign
galaxies still rely on uncertain empirical mass loss laws. With the advent of
upcoming new instrumentation we expect a much better understanding of the
processes near the photosphere. Monitoring will be mandatory in many cases to
describe correctly the effects of pulsation. The ambiguities due to uncertain
distances will be overcome by intensifying the study of AGB populations in the
Magellanic Clouds and extending studies towards other galaxies of the Local
Group.  However, due to the inherent variability of the stars, small sized
telescopes with robotic observing equipment will also be able to contribute
substantially to modern AGB research.

\bibliographystyle{aa}

\end{document}